\documentclass[epj]{svjour}
\usepackage{graphicx}

\begin{document}
\title{Direct photons in d+Au collisions at $\sqrt{s_{NN}}=200\,\mathrm{GeV}$ with STAR}

\author{M J Russcher for the STAR collaboration}

\institute{Utrecht University, Utrecht, The Netherlands \\
  \email{m.j.russcher@phys.uu.nl}}

\date{Received: date / Revised version: date}
% The correct dates will be entered by Springer

\abstract{Results are presented of an ongoing analysis of direct photon
production in $\sqrt{s_{NN}}=200\,\mathrm{GeV}$ deuteron-gold collisions (d+Au) with
the STAR experiment at RHIC. A significant excess of direct photons is
observed near mid-rapidity ($0<y<1$) and found to be consistent with 
next-to-leading order pQCD calculations including the contribution
from fragmentation photons.}


\maketitle

\section{Introduction}
\label{intro}
Direct photons are an interesting tool to study the quark-gluon plasma
(QGP) created in ultra-relativistic heavy-ion collisions \cite{PTphotons}. 
These photons are directly produced in processes like quark-anti-quark annihilation 
($q+\bar{q}\rightarrow g+\gamma$) and quark-gluon Compton scattering 
($q+g\rightarrow q+\gamma$) and do not originate from hadronic decays. 
Their main advantage is that they do not interact with the color charges when
traversing the dense medium that is formed in the heavy-ion collision. 

\begin{figure}[h]
  \centering
  \resizebox{0.45\textwidth}{!}{\includegraphics[angle=-90]{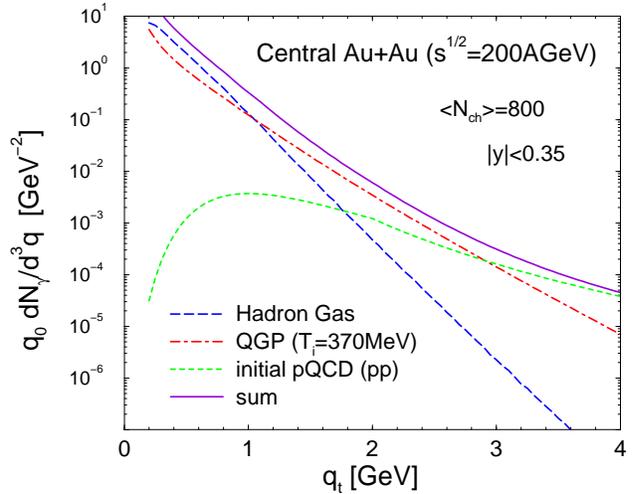}}
  \caption{Theoretical calculation for the mid-rapidity direct photon
    yield in central Au+Au collisions at RHIC as a function of 
    transverse momentum ($q_t$), coming from various
    sources. (figure from \cite{TRGphotons})}
  \label{fig:photons_theory}
\end{figure}

It is expected that thermal photons dominate the direct photon yield
at low transverse momentum ($p_T$). Since the thermal photon
spectrum falls off exponentially
with $p_T$, the prompt photons from the initial hard
(pQCD-like) scattering will dominate the spectrum at higher $p_T$,
as can be seen from the calculation shown in Figure \ref{fig:photons_theory}. 
In addition there is a contribution from photons produced
during the fragmentation of the partons. The yield of these
photons can be determined using parton-to-photon
fragmentation functions \cite{BFGfrag}.

The thermal photons are radiated by the electric char\-ges in the QGP 
and the hadron gas which is formed in the later stage of the
expansion. Their momentum distribution is therefore
a measure for the temperature of both phases \cite{TRGphotons}. 
A measurement of thermal photons can thus provide information on the temperature
evolution of the system.

The measurement of prompt photons is of interest, 
first of all because they form a
background to the measurement of the thermal component.
Furthermore, events with a tagged prompt photon and a 
recoil jet are a promising tool to study the interaction
of the jet with the medium \cite{XNgammajet}.

The production of high $p_T$ particles in nucleus-nucleus (A+A)
collisions is often represented by the nuclear modification factor
\begin{equation}
  \label{eq:3}
  R_{AA} = \frac{{\rm d}N^{AA}/{\rm d}p_T}
  {\langle N_{bin} \rangle {{\rm d}N}^{pp}/{\rm d}p_T}
\end{equation}
which is the yield of hadrons in A+A collisions relative to a scaled 
reference spectrum measured in proton-proton (p+p) collisions.
The scale factor $\langle N_{bin} \rangle$ is the number of binary
nucleon-nucleon collisions which can be calculated in the
framework of a Glauber-model. 
For sufficiently hard (high $p_T$) processes it is expected that
particle production scales with $\langle N_{bin} \rangle$.
However, a suppression of hadrons by a factor of 5 was
observed \cite{BRquenching,PHquenching,PHOquenching,STquenching}
and has been attributed to in-medium parton energy loss.

Prompt photons provide a way to test binary collision scaling 
since their production is not affected by the medium produced
in the final state of the interaction. A recent measurement at RHIC
indeed shows that high $p_T$ direct photon production 
in gold-gold (Au+Au) interactions is consistent with 
$\langle N_{bin} \rangle$ scaling ($R_{AA} \sim 1$) \cite{PHdirgamma}.

At RHIC there is the possibility to study direct photon production
not only in Au+Au, but also in p+p and d+Au collisions.  
Direct photon production in p+p collisions can serve as a
high precision test of pQCD, while in d+Au collisions it can be used 
to investigate nuclear effects such as 
the existence of a Color Glass Condensate \cite{ILMcgc,ILMcgc2} 
and multiple rescattering (Cronin enhancement) \cite{Cronin}.
However, the measurement of direct photons, be it in p+p, d+Au
or Au+Au collisions, will always be challenging because of the large
background of photons from hadronic decays.

\section{Experiment}
\label{sec:1}
The data presented here were taken with the STAR detector \cite{STAR}
in the $\sqrt{s_{NN}}=200\,\mathrm{GeV}$ d+Au run at RHIC. 
A minimum bias trigger was provided by requiring a signal over threshold 
in the Zero Degree Calorimeter (ZDC) in the Au beam direction 
(at negative pseudorapidity).

The energy of the photon showers was measured with the 
Barrel Electromagnetic Calorimeter (BEMC) \cite{BEMC} consisting 
of 4800 lead-scintillator cells with a spatial granularity 
of $\Delta \eta \times \Delta \phi = 0.05\times0.05$.
The BEMC is positioned at a distance of $2.3\,\mathrm{m}$ from the beam axis
and covers full azimuth in the pseudorapidity interval $|\eta|<1$. 
During the d+Au run however, only half of the detector $(0<\eta<1)$ was instrumented. 
The absolute energy scale of the BEMC was calibrated 
to a precision of $5\%$ using minimum ionizing particles and electrons
reconstructed in the Time Projection Chamber (TPC) \cite{TPC}.

A gaseous wire-proportional counter with strip readout is 
located inside the BEMC at a depth of 5--7 radiation lenghts \cite{BEMC}.  
The finer segmentation 
$(\Delta \eta \times \Delta \phi = 0.007\times0.007)$
of this Barrel Shower Maximum Detector (BSMD) allows to measure the
transverse profile of the showers and makes it possible to resolve
the two $\pi^{0}$ decay photons in the high $p_{T}$ region.  

To enhance the particle yield at high
$p_{T}$, a level-0 trigger selected events with a
high transverse energy deposition in a single BEMC cell.
The integrated luminosity, after all
event cuts, is $105~\mu\textrm{b}^{-1}$ and $996~\mu\textrm{b}^{-1}$
associated with an effective $p_T$ threshold set at 2.5 and 4.5~GeV,
respectively.

\section{Analysis}
\label{sec:2}
This analysis aims to measure the direct photon yield in d+Au collisions
by means of a statistical subtraction of the hadronic decay background 
from the measured inclusive photon spectrum. The dominant contribution 
to this background is from the decay $\pi^{0} \rightarrow \gamma\gamma$. 
Therefore it is important to constrain the $\pi^{0}$ yield with high precision. 
This yield has been determined in an earlier analysis of the present data
as reported in \cite{STpions}.

The total photon yield from hadronic decays, see Table~\ref{tab:decaytable}, 
was simulated with a Monte-Carlo algorithm. 
\begin{table}
  \caption{Dominant meson decay contributions to the inclusive photon yield.
    The factor used for scaling the transverse mass spectra
    (see text) is given in the last column.
  }
  \label{tab:decaytable}
  \begin{tabular}{lcc}
    \hline\noalign{\smallskip}
    decay & branching ratio & $m_T\mathrm{-scale}$ \\
    \noalign{\smallskip}\hline\noalign{\smallskip}
    $\pi^0 \rightarrow \gamma\gamma$  & 98.80\% & N/A \\
    $\pi^0 \rightarrow e^+e^-\gamma$  & 1.20\%  & \\
    \noalign{\smallskip}\hline\noalign{\smallskip}   
    $\eta \rightarrow \gamma\gamma$   & 39.23\%  & 0.45 \\
    $\eta \rightarrow \pi^+\pi^-\gamma$ & 4.78\% & \\
    $\eta \rightarrow e^+e^-\gamma$ & 0.49\% & \\
    \noalign{\smallskip}\hline\noalign{\smallskip}
    $\omega\mathrm{(782)} \rightarrow \pi^0\gamma$ & 8.69\% & 1.0 \\
    \noalign{\smallskip}\hline   
  \end{tabular}
\end{table}
Input to this simulation is a fit to the measured $\pi^0$ spectrum. 
Since there are no measurements available of the $\eta$ and $\omega$(782) yields, 
we assume that their transverse mass ($m_T$) spectra scale with that of the 
$\pi^0$ by a constant factor. 
In this analysis, these factors were taken to be $\eta/\pi^0 = 0.45 \pm 0.05$ 
and $\omega/\pi^0 = 1 \pm 0.2$, consistent with measurements reported
in \cite{PHetas} and \cite{ISRomegas}. The other possible contributions
to the hadronic decay background were found to be negligible. 
In figure 2 we show the $p_T$ dependence of the decay photons, 
normalized to the generated $\pi^0$ spectrum. 
\begin{figure}
\centering
  \resizebox{.45\textwidth}{!}{\includegraphics{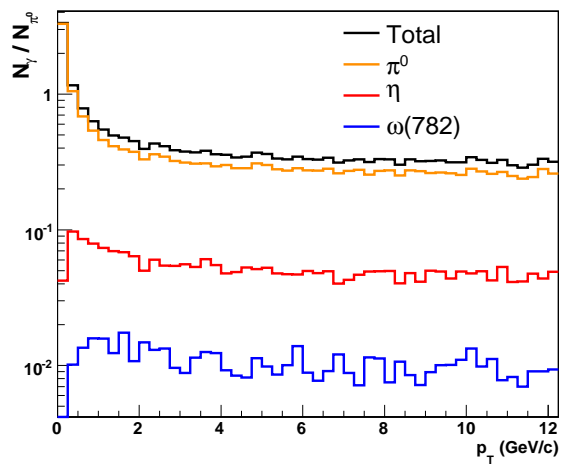}}
  \caption{The number of photons per input pion $N_{\gamma}/N_{\pi^0}$
    from hadronic decays versus $p_T$, obtained from a simulation 
    described in the text.}
\label{fig:bgphotons}
\end{figure} 

In this analysis, inclusive photon candidates were identified by a clustering 
algorithm based on the energy measured in the BEMC and on the shower 
profile measured in the BSMD. To identify neutral clusters and decrease
hadronic background, a charged particle veto 
(CPV) is provided by rejecting clusters with a pointing TPC track.

The raw inclusive photon yield has been corrected for (anti-)neutron contamination
by means of a GEANT simulation of the detector, which had the measured
proton and anti-proton spectra as an input \cite{STprotons}.
Correction factors to account for reconstruction and trigger efficiency, 
limited acceptance, photon conversions in the detector material
and the inefficiency of the CPV were determined from an analysis
of GEANT hits embedded in real d+Au data. 
\begin{figure}[h]
\centering
  \resizebox{.45\textwidth}{!}{\includegraphics{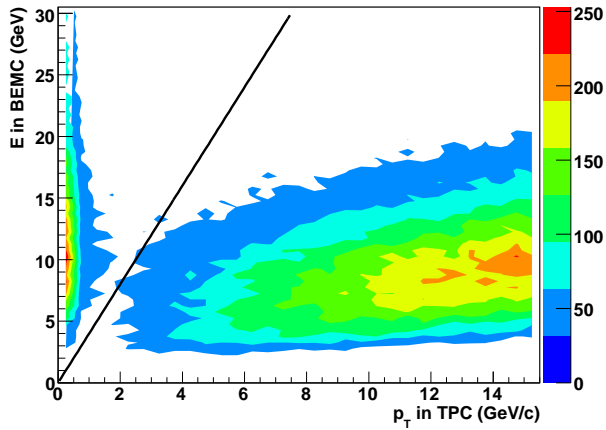}}
  \caption{The energy deposited in the BEMC versus the total momentum 
of charged tracks, reconstructed with the TPC. There is a sizeable
contribution of events which have a large amount of energy in the
BEMC but very little TPC momentum. The full line shows
the cut that was used to exclude these events from the analysis.}
\label{fig:bgratio} 
\end{figure}

A prominent background to the measurement is due to
scattering of the deuteron beam halo on material
located upstream of the STAR interaction region. Such events are
characterized by a large deposition of energy in the BEMC
without associated tracks reconstructed in the TPC and thus do 
have the signature of a collection of neutral clusters.
A cut on the ratio of the TPC momentum to the BEMC energy 
was used to remove these events from the 
data sample, see Figure \ref{fig:bgratio}.

\section{Results}
\label{sec:results}
To determine the direct photon yield we have calculated the double ratio
\begin{equation}
  \label{eq:1}
  R = \frac{(\gamma/\pi^{0})_{measured}}{(\gamma/\pi^{0})_{decay}} 
  = 1 + \frac{\gamma_{direct}}{\gamma_{decay}}
\end{equation}
where the numerator is the point by point ratio of the measured inclusive photon
spectrum to the neutral pion spectrum and the denominator is the number
of simulated decay photons per input pion as shown in Figure \ref{fig:bgphotons}. 
It is clear that many systematic uncertainties, which are common to neutral 
pion and inclusive photon detection will (partially) cancel in this ratio. 
The remaining uncertainties in the ratio $R$ are listed in Table \ref{tab:errortable}.
\begin{table}
  \caption{Dominant contributions to the systematic error on the 
    ratio $R$ defined in equation \ref{eq:1} for two $p_T$ bins at 3 and 
    10~GeV/$c$.}
  \label{tab:errortable}
  \begin{tabular}{lcc}
    \hline\noalign{\smallskip}
    $(\gamma/\pi^{0})_{meas} / (\gamma/\pi^{0})_{bg}$ & 
    $3\,\mathrm{GeV}/c$ & $10\,\mathrm{GeV}/c$\\
    \noalign{\smallskip}\hline\noalign{\smallskip}
    pion yield extraction & $5\%$ & $10\%$ \\
    reconstruction efficiency (stat.) & $10\%$ & $4.5\%$ \\
    BSMD gain uncertainty & $5\%$ & $5\%$ \\
    beam background & $<1\%$ & $4\%$ \\
    BEMC energy scale & $3\%$ & $3\%$ \\
    $\eta/\pi^{0}=0.45\pm0.05$ & $3\%$ & $3\%$ \\
    \noalign{\smallskip}\hline   
  \end{tabular}
\end{table}
\begin{figure}
  \centering
  \resizebox{.48\textwidth}{!}{\includegraphics{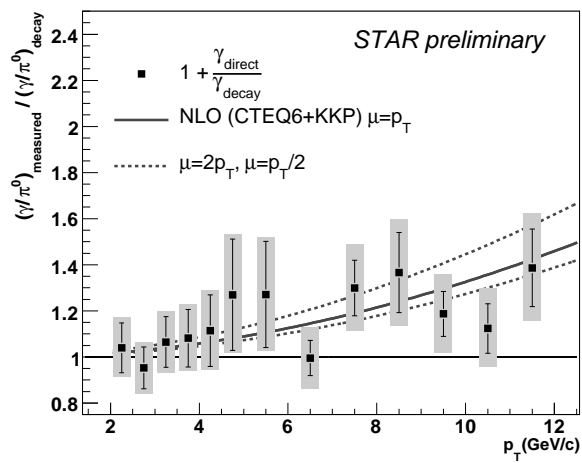}}
  \caption{The double ratio $R$, defined in equation \ref{eq:1}, as a function of $p_T$.
    The error bars (grey boxes) indicate the statistical (total) 
    error on the data points as specified in Table \ref{tab:errortable}.
    The full line is a pQCD calculation for p+p collisions using the
    CTEQ6M parton densities and KKP fragmentation functions. The dashed 
    lines show the sensitivity of the calculation to the factorization scale.}
  \label{fig:ratio_dA} 
\end{figure}

In Figure \ref{fig:ratio_dA} we show the $p_T$ dependence 
of $R$ obtained from the d+Au data set. 
The significant excess above unity at large $p_T$ indicates the presence 
of a direct photon signal.
The double ratio is consistent with a pQCD calculation \cite{pQCDvogel} based
on the CTEQ6M parton density functions \cite{CTEQ6m} 
and KKP fragmentation functions \cite{KKP} as
shown for three different factorization scales by the curves
in Figure \ref{fig:ratio_dA}. Since this analysis does not make use of an
isolation cut the calculated direct photon yield includes both prompt
and fragmentation photons.

\section{Outlook}
\label{sec:outlook}
The significance of the measurement will be much improved by reducing
the error contribution from the BSMD which has not been calibrated in-situ so far.
In addition, dedicated TPC tracking algorithms
and BEMC pattern recognition algorithms are presently being developed 
to better identify beam-background events which will reduce the
systematic uncertainty from this source and the statistical error on the 
efficiency correction is also being improved.

The direct photon analysis of the 2005 p+p data set is in progress. 
The combined d+Au and p+p results will provide insight into nuclear 
effects and constitute a necessary baseline to study the properties 
of the QGP from direct photon measurements in Au+Au collisions.


\begin{thebibliography}{99}
\bibitem{PTphotons} Peitzmann T and Thoma M H 2002 \textit{Phys. Rept.} {\bf 346} 175
\bibitem{TRGphotons} Turbide S \textit{et al} 2004 \textit{Phys. Rev.} C {\bf 69} 014903
\bibitem{BFGfrag} Bourhis L \textit{et al} 1998 \textit{Eur.Phys.J.} C {\bf 2} 529
\bibitem{XNgammajet} Wang X N \textit{et al} 1996 \textit{Phys. Rev. Lett.} {\bf 77} 231
\bibitem{BRquenching} Arsene I \textit{et al} 2003 \textit{Phys. Rev. Lett.} {\bf 91} 072305
\bibitem{PHquenching} Adler S S \textit{et al} 2003 \textit{Phys. Rev. Lett.} {\bf 91} 072303
\bibitem{PHOquenching} Back B B \textit{et al} 2003 \textit{Phys. Rev. Lett.} {\bf 91} 072302
\bibitem{STquenching} Adams J \textit{et al} 2003 \textit{Phys. Rev. Lett.} {\bf 91} 072304
\bibitem{PHdirgamma} Adler S S \textit{et al} 2005 \textit{Phys. Rev. Lett.} {\bf 91} 232301
\bibitem{ILMcgc} Iancu I \textit{et al} 2001 \textit{Nucl. Phys.} A {\bf 692} 583
\bibitem{ILMcgc2} Ferreiro E \textit{et al} 2002 \textit{Nucl. Phys.} A {\bf 703} 489
\bibitem{Cronin} Cronin J \textit{et al} 1975 \textit{Phys. Rev.} D {\bf 11} 3105
\bibitem{STAR} Ackermann K H \textit{et al} 2003 \textit{Nucl. Instrum. Meth.} A {\bf 499} 624
\bibitem{BEMC} Beddo M \textit{et al} 2003 \textit{Nucl. Instrum. Meth.} A {\bf 499} 725
\bibitem{TPC} Anderson M \textit{et al} 2003 \textit{Nucl. Instrum. Meth.} A {\bf 499} 659
\bibitem{STpions} Mischke A \textit{et al} 2005 \textit{Eur. Phys. J.} C {\bf 43} 311  
\bibitem{PHetas} Adler S S \textit{et al} 2006 \textit{Phys. Rev. Lett.} {\bf 96} 202301 
\bibitem{ISRomegas} Diakonou M \textit{et al} 1980 \textit{Phys. Lett.} B {\bf 89} 432
\bibitem{STprotons} Adams J \textit{et al} 2006 \textit{Phys. Lett.} B {\bf 637} 161
\bibitem{pQCDvogel} Vogelsang W 2006 \textit{private communication}
\bibitem{CTEQ6m} Pumplin J \textit{et al} 2002 \textit{J. High Energy Phys.} {\bf 07} 012
\bibitem{KKP} Kniehl B A \textit{et al} 2001 \textit{Nucl. Phys.} B {\bf 597} 337
\end{thebibliography}
\end{document}